\documentclass[conference]{IEEEtran}
\IEEEoverridecommandlockouts
\usepackage{cite}
\usepackage{amsmath,amssymb,amsfonts}
\usepackage{bbm}
\usepackage{algorithm,algorithmic}
\usepackage{graphicx}
\usepackage{textcomp}
\usepackage{xcolor}
\def\BibTeX{{\rm B\kern-.05em{\sc i\kern-.025em b}\kern-.08em
    T\kern-.1667em\lower.7ex\hbox{E}\kern-.125emX}}
\usepackage{float}
\usepackage{caption}
\usepackage{cases}
\usepackage{stfloats}
\newtheorem{theorem}{Theorem}
\newtheorem{lemma}{Lemma}

\newtheorem{corollary}{Corollary}
\newtheorem{pb}{Problem}

\graphicspath{{figures/}}
\allowdisplaybreaks
\newcommand{\haoyue}{\color{black}}

\begin{document}

\title{Delay Optimal Cross-Layer Scheduling Over Markov Channels with Power Constraint}
\author{
	\IEEEauthorblockN{Wenhao~Zhan\textsuperscript{1},~Haoyue~Tang\textsuperscript{1},~Jintao~Wang\textsuperscript{1,2}}
	\IEEEauthorblockA{
		\textsuperscript{1}Beijing National Research Center for Information Science and Technology (BNRist),\\
		Dept. of Electronic Engineering, Tsinghua University, Beijing 100084, China\\
		\textsuperscript{2}Research Institute of Tsinghua University in Shenzhen, Shenzhen, 518057\\
		\{zhanwh17@mails, thy17@mails, wangjintao@\}tsinghua.edu.cn}}

\maketitle

\begin{abstract}
	We consider a scenario where {\haoyue a power constrained transmitter delivers randomly arriving packets to the destination }over Markov time-varying channel and adapts different transmission power to each channel state in order to guarantee successful transmission. To {\haoyue minimize the expected average transmission delay of each packet}, we formulate the problem into a constrained Markov decision process (CMDP). We reveal the queue-length threshold structure of the optimal policy, i.e., the transmitter sends packets if and only if the queue length surpasses a threshold and obtain the optimal cross-layer scheduling strategy through linear programming (LP). Numerical results {\haoyue validate the performance of the proposed strategy and illustrate} a delay-power tradeoff in such scenario. 
\end{abstract}

\begin{IEEEkeywords}
Cross-layer Control, delay-power trade-off, Constrained Markov Decision Process (CMDP).
\end{IEEEkeywords}

\section{Introduction}
\let\thefootnote\relax\footnotetext{\noindent -----------------\\ This work was supported in part by the National Key R\&D Program of China under Grant 2017YFE0112300, Beijing National Research Center for Information Science and Technology under Grant BNR2019RC01014 and BNR2019TD01001. Corresponding author: Jintao Wang. }

The proliferation of the Internet of Things (IoT) network and real time services raises high reliable and low latency data communication requirements on future networks. Due to the large number of access nodes, each communication device in such network is  equipped with limited power. To meet both low latency and high throughput requirements under power constraint in time-varying wireless networks, efficient transmission strategy is needed. 

Cross-layer control strategy has been an effective approach to reduce transmission delay in time-varying channels with limited power \cite{wang_17_GC,wang_19_tcomm,yang_10_twc,uysal_02_ton,berry_02_tit,haoyue}. By classifying channel states into "\emph{Good}" and "\emph{Bad}", previous work \cite{uysal_02_ton,berry_02_tit} studied energy efficient transmission strategy to minimize queueing delay for a point to point communication system. When different levels of power can be used to guarantee successful transmission in different channel states, Wang \emph{et al.} derived the optimal stationary transmission policy that minimizes the queueing delay under an average power constraint \cite{wang_17_GC,wang_19_tcomm}. Notice that the above work assumed the channel fading to be an i.i.d process. A more practical assumption to model the time-varying channel is to assume that channel states evolve as a Markov chain similar to \cite{chen_2018_ton,tang2019minimizing}.

To characterize the delay-power trade-off under a more practical Markov time-varying channel, we study joint queue aware and channel aware cross-layer control strategy to obtain a minimum delay performance. We formulate the delay-optimal scheduling problem into a constrained Markov decision process and obtain the optimal stationary randomized policy through Linear Programming (LP). Finally, the performance of the proposed algorithm is evaluated through simulations.

The rest of the paper is organized as follows: we formulate the overall scheduling problem in Section II. In Section III, we formulate the scheduling problem into a constrained Markov decision process (CMDP) and analyze its optimal structure. The optimal solution to the CMDP is then obtained through Linear Programming (LP). Section IV provides simulation results and Section V draws the conclusion.

\emph{Notations: }The probability of event $A$ is denoted as $\text{Pr}\{A\}$ and the expectation is denoted by $\mathbb{E}[A]$. 
\section{System Model}

\subsection{Network Model}

We consider the scheduling policy for {\haoyue a discrete time point-to-point communication system} and let $n\in\mathbb{N}^+$ denote the index of slots. The number of packets arriving at {\haoyue the transmitter} in the $n$-th slot, denoted by {\haoyue $b[n]\in\{0, 1\}$}, follows an i.i.d Bernoulli distribution with expectation $\mathbb{E}[b[n]]=\theta$.
%
{\haoyue Those packets wait in an FIFO queue at the transmitter before they are sent out to the receiver and the queue length in slot $n$ is captured by $l[n]$. We assume the undelivered packets wait in a finite buffer of size $K$ and when the current queue length is about to exceed the buffer capacity, i.e, $l[n]=K$, arriving packets in th next slot will be discarded. } 

{\haoyue We assume packets are sent from the transmitter to the receiver through a time-varying wireless link. }Assume that the channel state remains invariant in each slot and we model the channel $s[n]$ in each slot as an ergodic $S$-state Markov chain. Let $P_{ij}$ denotes the conditional probability that channel state evolves from state $i$ to state $j$, i.e., 
\begin{equation}
\text{Pr}\{s[n]=j|s[n-1]=i\}\triangleq P_{ij}, i,j=1,2,...,S.
\end{equation}

With no loss of generality, we assume large $s[n]$ indicates better channel quality and thus less power is needed to transmit a packet. When the current channels state is $s$, i.e., $s[n]=s$, the transmitter uses $X_s$ power to guarantee successful transmission in channel state $s$. Thus $
X_s>X_{s'}, \forall s<s'$. In this work, similar to \cite{wang_17_GC}, we assume that at most one packet can be transmitted in each time slot and the transmitted packet will be successfully received at the end of the slot. Let $a[n]\in\{0, 1\}$ denote the decision in $n$-th slot, where $a[n]=1$ denotes a transmission decision is made and $a[n]=0$ indicates that the channel remains idle. To minimize the queueing delay under an average power constraint, the transmission decision $a[n]$ is made based on the current channel state $s[n]$ and queue length $q[n]$. 
The average power consumed over consecutive $N$ slots can thus be characterized by:
\begin{equation}    
E_N=\frac{1}{N}\sum_{n=1}^Na[n]X_{s[n]}              
\end{equation}

And the queue length evolution is as follows:
\begin{equation}    
l[n] = \max\{\min\{l[n-1]+b[n],K\}-a[n],0\}.           
\end{equation}

Here we assume the buffer size $K$ is large enough so that packet-loss due to full buffer is unlikely to happen. Then the average transmission delay of each packet can be computed by:
\[\lim_{N\to\infty}\frac{1}{N}\sum_{n=1}^{N}l[n].\]

\subsection{Problem Formulation}
Our goal is to design a non-anticipated policy $\pi$ that utilizes the current queue length and channel state information in making schedule decisions to minimize the average queueing delay under power constraint. Denote $\pi_{\text{NA}}$ to be the set of non-anticipated scheduling decisions, then the optimization problem is organized as follows:
\begin{pb}[Delay Optimal Scheduling Problem]
\begin{subequations}

\begin{equation}    
\pi^*=\arg\min_{\pi\in\Pi_{\text{NA}}}D(\pi), \text{where }D(\pi) = \lim_{N\to\infty}\mathbb{E}_{\pi}\left[\frac{1}{N}\sum_{n=1}^Nl[n]\right] ,
\end{equation}
\begin{equation}
	\text{s.t. }\lim_{N\to\infty}\mathbb{E}_\pi\left[\frac{1}{N}\sum_{n=1}^Na[n]X_{s[n]}\right]\leq\mathcal{E}.
\end{equation}
\end{subequations}
\end{pb}

\section{Problem Resolution}
In this section we first {\haoyue formulate} the problem as a constrained Markov decision process (CMDP) and verify the threshold structure of the optimal policy. Then we derive the solution through linear 
programming (LP).
\subsection{Constrained Markov Decision Process Formulation}
The scheduling problem can be formulated into a CMDP with the following four parts:
\begin{itemize}
	\item [(1)] 
	\textbf{State Space:} The state in slot $n$ can be characterized by the queue length and channel state $(l[n],s[n])$.
	\item [(2)]
	\textbf{Action Space:} The source node can choose two possible actions in each slot. Action $a[n]=1$ denotes that the node schedules to transmit a packet, while $a[n]=0$ means that the transmitter stays idle in the slot. Thus the action space $\mathbb{A}=\{0,1\}$. 
	\item [(3)]
	\textbf{Transfer Function:} The queue length in the next slot relies on the number of arriving packets and scheduling decision. If $a[n]=1$, then $q[n+1]=q[n]+b[n]-1$. Otherwise $q[n+1]=q[n]+b[n]$. Since the channel state evolves independently, the transfer function can be computed as follows:
    \begin{subequations}
    \begin{align}
	&\text{Pr}\{(q,s)\rightarrow (q+1,s')\}=P_{ss'}\theta,\quad a=0;\\	
	&\text{Pr}\{(q,s)\rightarrow (q,s')\}=P_{ss'}(1-\theta),\quad a=0;\\
	&\text{Pr}\{(q,s)\rightarrow (q,s')\}=P_{ss'}\theta,\quad a=1;\\
    &\text{Pr}\{(q,s)\rightarrow (q-1,s')\}=P_{ss'}(1-\theta),\quad a=1.
	\end{align}
    \end{subequations}
    \item [(4)]\textbf{One-Step Cost:} The one-step cost includes delay cost and power cost. Let $C_{Q}(q,s,a)$ denote the delay cost while $C_{X}(q,s,a)$ the power cost, then   
    \begin{subequations}
    	\begin{equation}    
        C_{Q}(q,s,a)=q,
    	\end{equation}
    	\begin{equation}
    	C_{X}(q,s,a)=X_{s}a.
    	\end{equation}
    \end{subequations}
\end{itemize}

With the introduction of the above four elements, the Delay Optimal Scheduling Problem can be expressed as the following finite state constrained Markov Decision Process (CMDP):
\begin{subequations}
	\begin{equation}  
	\pi^*=\arg\min_{\pi\in\Pi_{\text{NA}}}D(\pi),
	\end{equation}
	\begin{equation}
    \text{where }D(\pi) = \lim_{N\to\infty}\frac{1}{N}\mathbb{E}_{\pi}\left[\sum_{n=1}^NC_{Q}(l[n],s[n],a[n])\right] ,
	\end{equation}
	\begin{equation}
	\text{s.t. }\lim_{N\rightarrow\infty}\frac{1}{N}\mathbb{E}_\pi\left[\sum_{n=1}^NC_{X}(l[n],s[n],a[n])\right]\leq\mathcal{E}.
	\end{equation}
\end{subequations}

\subsection{Queue-Length Threshold Structure}
In this part we study the structure of the delay optimal policy. First we introduce a corollary from \cite{altman1999constrained} to illustrate a basic property of the policy. Here deterministic policies refer to those whose transmission probability $f_{q,s}$ is either 1 or 0.
\begin{corollary}
	An optimal stationary policy $\pi^*$ is a mixture of two deterministic policies $\pi_{1},\pi_{2}$. Let $\lambda$ denote the weight of $\pi_{1}$, then in each slot $n$, policy $\pi^*$ selects policy $\pi_1$ with probability $\lambda$ and policy $\pi_2$ with probability $1-\lambda$, i.e.,
	\begin{equation}
	\pi^*=\lambda \pi_{1}+ (1-\lambda)\pi_{2}.
	\end{equation}
\end{corollary} 

{\haoyue To obtain $\pi^*$ through LP, first we need to know the structure of $\pi^*$. This optimum structure is obtained by analyzing the structure of $\pi_1$ and $\pi_2$.} To obtain $\pi_1$ and $\pi_2$, we place the power constraint into the objective function with the Lagrangian multiplier $\eta\geq0$: \begin{equation}
\label{MDP}
\begin{split}&\min_{\pi\in\Pi_{\text{NA}}}\lim_{N\to\infty}\frac{1}{N}\mathbb{E}_{\pi}\Bigg[\sum_{n=1}^NC_{Q}(l[n],s[n],a[n])+ \\ 
&\hspace{1.5cm}\eta(C_{X}(l[n],s[n],a[n])-\mathcal{E})\Bigg].
\end{split}
\end{equation} 

%

\begin{theorem}
\label{threshold structure}
	Given $\eta$, the optimal deterministic stationary policy to the unconstrained scheduling problem (\ref{MDP}) is controlled by a queue length threshold $L_{s}$ for each channel state $s$. When $l[n-1]+b[n]\geq L_{s}$, the optimum policy transmits a packet i.e., $a[n]=1$, otherwise the transmitter idles and $a[n]=0$.
\end{theorem}
\begin{IEEEproof}
	The detail is provided in Appendix A.
\end{IEEEproof}

Theorem~\ref{threshold structure} indicates following the optimum strategy we will have a finite queue length, which is shown in the corollary below:
\begin{corollary}
	When the queue capacity $K$ is sufficiently large, following the optimum stationary policy $\pi^*$, the queue length $l[n]$ will never reach $K$.
\end{corollary}
\begin{IEEEproof}
	Suppose that the threshold values of the optimal deterministic policy $\pi_{1}$ and $\pi_{2}$ are $L_{1}^{1},L_{2}^{1},...,L_{S}^{1}$ and $L_{1}^{2},L_{2}^{2},...,L_{S}^{2}$, respectively. Denote $L^{*}=\max_{i,s}L_{s}^{i}.$
	According to Theorem~\ref{threshold structure}, since the optimum stationary policy $\pi^*$ is a mixture of $\pi_1$ and $\pi_2$, whenever the queue length reaches $L^{*}$, policy $\pi^*$ will transmit a packet, and thus the queue length will not exceeds $L^*$. As long as the $K$ is selected sufficiently large, i.e., $K\ge L^{*}+1$, we can guarantee the queue length $l[n]$ following policy $\pi^*$ is smaller than $K$.  
\end{IEEEproof}

\subsection{Linear Programming Formulation}
According to the previous analysis, the optimum policy to the above CMDP has a threshold structure. In this section, we obtain the optimum policy through LP. 

Recall that we assume the buffer size $K$ is large enough so that the queue length is always smaller than $K$. For each stationary policy $\pi$, denote $\mu_{q,s}$ as the steady state distribution that the current queue length is $q$ and the channel state is $s$. Let $y_{q,s}$ be the probability that the current queue length is $q$, the channel state is $s$ and the transmitter schedules to send one packet. To transform the CMDP into an equivalent LP, first we introduce a set of variables $y_{q,s}=\mu_{q+1,s}(1-\theta)f_{q+1,s}+\mu_{q,s}\theta f_{q+1,s}$, which denotes the probability of having $q$ packets waiting in the queue after sending a packet in channel state $s$. The following theorem helps us to transform the CMDP into an equivalent LP:
\begin{theorem}
	\label{LP}
	Solving the Delay-Opt Stationary Problem is equivalent to solve the following LP problem:
	\begin{subequations}
		\begin{equation}
		\label{delay}
		\text{D}^{\text{opt}}=\min_{\{y_{q,s}\}}\frac{1}{\theta^{2}}\sum_{q=0}^K\sum_{s=1}^Sqy_{q, s},
		\end{equation}
    \begin{align}
    \text{s.t. }\quad &\label{power}\sum_{q=0}^{K}\sum_{s=1}^{S}X_{s}y_{q,s}\leq\mathcal{E},\\
    &\label{packets}\sum_{q=0}^{K}\sum_{s=1}^{S}y_{q,s}=\theta,\\
    &\label{channel}\sum_{q=0}^{K}\boldsymbol{g_{q,s}^{T}Y}=\rho_{s}, \forall S,\\
    &\label{probability}0\leq y_{q,s}\leq(1-\theta)\boldsymbol{g_{q,s}^{T}Y}+\theta\boldsymbol{g_{q,s}^{T}Y},\forall q,s,\\
    &\label{state}0\leq\boldsymbol{g_{q,s}^TY}\leq 1,\forall q,s.
    \end{align}
	\end{subequations}
where $\boldsymbol{Y}=[y_{0,1},y_{1,1},y_{2,1},...,y_{K,1},y_{0,2},...,y_{K,S}]^T$ and $\rho_{s}$ is the stationary distribution of $\mathbf{P}$. $\mathbf{G}=[\boldsymbol{g_{0,1},g_{1,1},g_{2,1},...,g_{K,1},g_{0,2},...,g_{K,S}}]^T$ is a matrix that characterizes the transformation from $\boldsymbol{Y}$ to $\boldsymbol{\mu}=[\mu_{0,1},\mu_{1,1},\mu_{2,1},...,\mu_{K,1},\mu_{0,2},...,\mu_{K,S}]^T$, i.e.,
\begin{equation}
\label{relation}
\boldsymbol{\mu}=\mathbf{G}\boldsymbol{Y}.
\end{equation}
$\mathbf{G}$ can be obtained from the following equations:
\begin{equation}
\begin{split}
\label{relationship}
&\sum_{s=1}^{S}P_{ss'}[(\mu_{q,s}\theta+\sum_{i=q+1}^{K}\mu_{i,s})-y_{q,s}]\\
&=\sum_{i=q+1}^{K}\mu_{i,s'}, \forall q,s.
\end{split}	
\end{equation}
\end{theorem}
\begin{IEEEproof}
	The detail is given in Appendix C.
\end{IEEEproof}

According to the above theorem, we present the power-delay tradeoff as the following corollary:
\begin{corollary}
	The optimal delay $\text{D}^{\text{opt}}$ monotonically decreases with the available power $\mathcal{E}$. 
\end{corollary}
\begin{IEEEproof}
	Suppose that $\text{D}^{\text{opt}}(\mathcal{E})=D$ when the power constraint is $\mathcal{E}$. Now we consider $\mathcal{E'}>\mathcal{E}$. Denote $y_{k}=\sum_{s=1}^{S}y_{k,s}$, then (\ref{delay}),(\ref{packets}) are equivalent to:
	\begin{equation}
	\label{newdelay}
	D=\frac{1}{\theta^{2}}\sum_{q=0}^Kqy_{q},
	\end{equation}
	\begin{equation}
	\label{newpackets}
	\sum_{q=0}^Ky_{q}=\theta
	\end{equation}
	Since $\mathcal{E'}>\mathcal{E}$, there exists $q_0$ so that we can construct the following new policy:
	\begin{subequations}
		\begin{equation}
		y'_{q}=y_{q}+\delta y_{q}, q\leq q_0,
		\end{equation}
		\begin{equation}
        y'_{q}=y_{q}-\delta y_{q}, q>q_0.
        \end{equation}
        where $\delta y_{q}\geq0$.		
	\end{subequations}
    We can choose sufficient small $\delta y_{q}$ such that $\sum_{q=0}^{q_0}X_{1}\delta y_{q}\leq \mathcal{E'}-\mathcal{E}$. Then
    \begin{equation*}
    \begin{split} 
    &\quad\sum_{q=0}^{K}\sum_{s=1}^{S}X_{s}y'_{q,s}\\
    &\leq\sum_{q=0}^{K}\sum_{s=1}^{S}X_{s}y_{q,s}+\sum_{q=0}^{q_0}\sum_{s=1}^{S}X_{s}\delta y_{q,s}\\
    &\leq\sum_{q=0}^{K}\sum_{s=1}^{S}X_{s}y_{q,s}+\sum_{q=0}^{q_0}X_{1}\delta y_{q}\\
    &\leq\mathcal{E'}.
    \end{split}
    \end{equation*}
    Notice (\ref{channel}) is only related to the sum of all queue length under one particular channel state. Hence, given $y'_{q}$, we can always adjust the values of $y'_{q,s}$ so that (\ref{channel}) is satisfied. From (\ref{newpackets}), it can be easily observed that $\sum_{q=0}^{q_{0}}\delta y_{q}=\sum_{q=q_{0}+1}^{K}\delta y_{q}$. Thus,
    \begin{equation*}
    \begin{split}
    D'-D&=\frac{1}{\theta^{2}}\sum_{q=0}^K(y'_{q}-y_{q})\\
    &\leq\frac{1}{\theta^{2}}(q_{0}\sum_{q=0}^{q_{0}}\delta y_{q}-(q_{0}+1)\sum_{q_{0}+1}^{K}\delta
y_{q}\\    
    &\leq 0.
    \end{split}
    \end{equation*}
    Therefore $\text{D}^{\text{opt}}(\mathcal{E'})\leq D'\leq D$.
\end{IEEEproof}

\section{Simulations}
In this part simulation results are first provided to validate our proposed optimal policy and the trade-off between power and delay is revealed. We assume a 3-state channel which satisfies the majorization relationship and the transmission matrix is as follows:
 \begin{equation}    
\mathbf{P}=
\begin{bmatrix}
0.5&0.3&0.2\\0.3&0.4&0.3\\0.2&0.3&0.5
\end{bmatrix}          
 \end{equation}
 The power consumed in each channel state is $\{X_1,X_2,X_3\}=\{4.5,1.5,0.5\}$
 and packets arrive at a rate of $\theta=0.6$. Queue capacity is constrained by $K=11$. We simulate each setting over $10^6$ slots. We compare the proposed scheduling policy with the greedy policy that transmits a packet whenever the transmitter has enough power. 
 
\begin{figure}[h]
	\centering
	\includegraphics[width=0.5\textwidth]{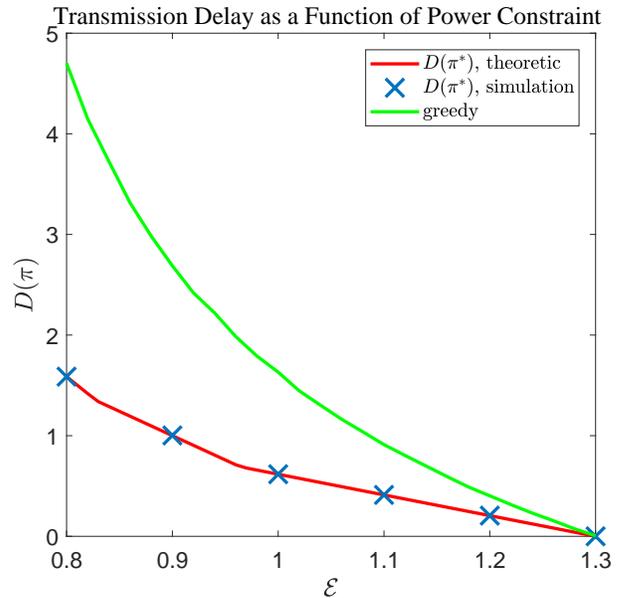}
	\caption{Average delay of optimal policies and greedy policies under different power constraint}
	\label{fig:performance}
\end{figure}

We first evaluate the delay performance of our proposed policy. Fig.~\ref{fig:performance} plots the average delay of the optimal scheduling policy and greedy policy with power constraint $\mathcal{E}\in[0.8,1.3]$. Moreover, the performance of the optimal policy is much better than the greedy policy when available power is limited, which proves the effectiveness of our strategy. Notice that the difference between these two policies vanishes as the power supply increases. This indicates that the optimal policy behaves similarly to the greedy policy to achieve shorter delay when the power supply is sufficient.     

\begin{figure}[h]
	\centering
	\includegraphics[width=0.5\textwidth]{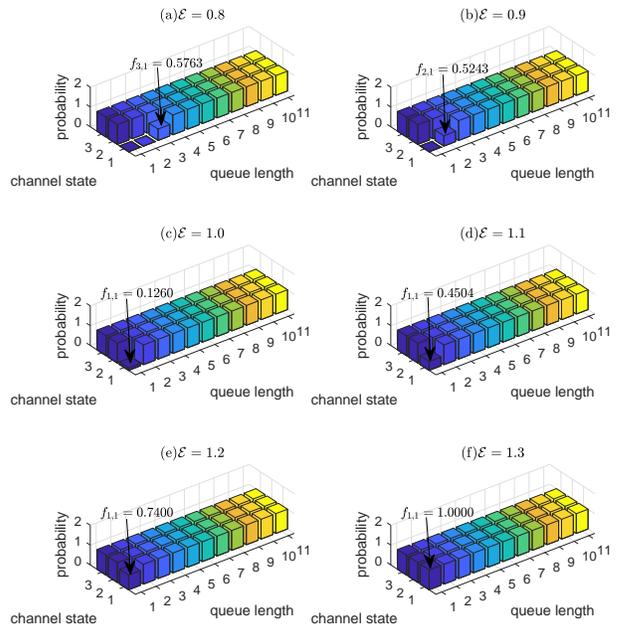}
	\caption{Threshold structure of optimal policies under different power constraint.}
	\label{fig:threshold}
\end{figure}

To study how our algorithm achieves such superior performance, Fig.~\ref{fig:threshold} displays the threshold structure of the optimal policy given different power constraint. When the available power increases, the threshold becomes smaller, which means that the source node transmits a packet more frequently. Besides, it is observed that the delay-power trade-off curve consists of three linear segments, each corresponding to a different threshold value of the policy. This verifies the linear property of the problem. 


\section{Conclusion}
In this paper we study the delay minimization scheduling problem under power constraint in wireless networks with ergodic Markov channels. We first reveal the threshold structure and then obtain the optimal policy through linear programming. It is shown that the optimal scheduling awaits and utilizes better channel quality while maintaining a small delay. In the future we will extend the work to more general scenarios with multi-user scheduling and bandwidth constraint.     
\bibliography{bibfile}
\begin{appendices} 
	\section{Proof of Theorem \ref{threshold structure}}
	The structure that minimizes the time-average cost of the Lagrange function \eqref{MDP} is obtained by the optimum structure that minimizes the $\alpha$-discounted cost and letting $\alpha\rightarrow1$. Following policy $\pi$, the $\alpha$-discounted cost starting from the state $(q,s)$, denoted by $J_{\alpha, \pi}(q, s)$ can be computed by:
	\begin{equation}
	\begin{split}
	&J_{\alpha,\pi}(q,s)=\lim_{N\to\infty}\mathbb{E}_{\pi}\Bigg[\sum_{n=1}^N\alpha^{n}[C_{Q}(l[n],s[n],a[n])+ \\ 
	&\hspace{0.5cm}\eta(C_{X}(l[n],s[n],a[n])-\mathcal{E})]|(l[0]=q,s[0]=s)\Bigg].
    \end{split}
	\end{equation} 
	Then, suppose $\pi_\alpha^*$ is the optimum policy that minimizes the $\alpha$-discounted cost. Then the value function $V_{\alpha}(q,s)=\min_{\pi}J_{\alpha, \pi}(q, s)$ satisfies the following Bellman equation:
	\begin{equation}
	\label{iteration}
	\begin{split}
	&V_{\alpha}(q,s)=\min\{C_Q(q,s,0)+\alpha\sum_{s'=1}^SP_{ss'}(\theta V_{\alpha}(q+1,s')\\
	&\hspace{1cm}+(1-\theta)V_{\alpha}(q,s')),C_Q(q,s,1)+\eta C_X(q,s,1)\\
	&\hspace{1cm}+\alpha\sum_{s'=1}^SP_{ss'}(\theta V_{\alpha}(q,s')+(1-\theta)V_{\alpha}(q-1,s'))\}
	\end{split}
	\end{equation}
	To validate the threshold structure, we first introduce the following lemma, proof details are similar to \cite[Lemma 3]{tang_twc}:
	\begin{lemma}
    \label{monotone}
		Fix discounted factor $\alpha$ and channel state $s$, then the difference of value function $d_{\alpha}(q,\cdot)=V_{\alpha}(q,\cdot)-V_{\alpha}(q-1,\cdot),q\geq1$ monotonically increases with $q$.
	\end{lemma}
	
Denote $\Delta V_{\alpha}(q,s)$ to be the difference of the sum of reward and the value function in the next state of taking $a_{\alpha}(q,s)=\{1,0\}$, i.e., 
	\begin{equation}
\begin{split}
\Delta V_{\alpha}(q,s)=&C_Q(q,s,1)+\eta C_X(q,s,1)\\
&+\alpha\sum_{s'=1}^SP_{ss'}(\theta V_{\alpha}(q,s')+(1-\theta)V_{\alpha}(q-1,s'))\\
&-C_Q(q,s,0)-\alpha\sum_{s'=1}^SP_{ss'}(\theta V_{\alpha}(q+1,s')\\
&-(1-\theta)V_{\alpha}(q,s')).
\end{split}
\end{equation}

Plugging $d_\alpha(q, \cdot)=V_\alpha(q, \cdot)-V_\alpha(q-1, \cdot)$ into the above equation we then have:
	\begin{equation}
	\begin{split}
	\Delta V_{\alpha}(q,s)=&\eta X_{s}-\alpha\sum_{s'=1}^{S}P_{ss'}(\theta d_{\alpha}(q+1,s')\\
	&+(1-\theta)d_{\alpha}(q,s')).
	\end{split}
	\end{equation}
	
	Since we have shown that $d_\alpha(q, \cdot)$ increases monotonically, it can be proved that $\Delta V_\alpha(q, \cdot)$ decreases monotonically. As a result, when $\Delta V_\alpha(q, s)<0$, it is indicated that the optimum policy $\pi_\alpha^*$ assigns an update decision in state $(q,s)$. Then for states $q'\geq q$, the optimum chooses to transmit because $\Delta V_\alpha(q', s)\leq\Delta V_\alpha(q, s)\leq 0$. Hence for each Lagrange multiplier $\eta$, the optimal policy $\pi_\alpha^*$ that minimizes the $\alpha$-discouted cost has two solutions: 
	(1) There exists a set of thresholds $L_s$. When the channel state is $s$, policy $\pi_\alpha^*$ sends a packet if the queue length surpasses the threshold $L_{s}$. 
	(2) Policy $\pi_\alpha^*$ never sends a packet. 
	
	Next we show the latter scenario is impossible. Assume that the source node never transmits a packet in any channel state and any queue length, then $\Delta V_{\alpha}(q,s)>0, \forall s, q$, which means that sequence $d_{\alpha}(q, s)$ is bounded for any $s$. According to Lemma~\ref{monotone}, we obtain that the limit point of the sequence exists. Denote that $\lim_{q\to\infty}d_{\alpha}(q,s')=\overline{d}_{s'}, \forall 1\leq s'\leq S$. Let $q\to\infty$, then for channel state $s_{1},s_{2},...,s_{r}$, the channel stays idle; for other states, the source node transmits a packet. For the first case we have:
    \begin{equation}
    \begin{split}
    &d_{\alpha}(q,s)=1+\alpha\sum_{s'=1}^{S}P_{ss'}(\theta d_{\alpha}(q+1,s')+\\
    &(1-\theta)d_{\alpha}(q,s')).
    \end{split}
    \end{equation}    
    Similarly for the second case we have:
    \begin{equation}
    \begin{split}
    & d_{\alpha}(q,s)=1+\alpha\sum_{s'=1}^{S}P_{ss'}(\theta d_{\alpha}(q,s')+\\
    &(1-\theta) d_{\alpha}(q-1,s')).
    \end{split}
    \end{equation}   
    Take $q\to\infty$, then in both cases the following equation holds:
    \begin{equation}
    \overline{d}_{s}=1+\alpha\sum_{s'=1}^{S}P_{ss'}\overline{d}_{s'}, \forall s.
    \end{equation}  
    Let $\alpha\to1$. Multiply each equation above by $\rho_{s}$ and sum them up, then we have:
    \begin{equation*}
    \sum_{s=1}^{S}\rho_{s}\overline{d}_{s}=1+\sum_{s=1}^{S}\rho_s\overline{d}_{s}
    \end{equation*}
    which clearly contradicts itself. Thus the threshold for each channel state exits, which verifies the structure of the optimal policy in Theorem~\ref{threshold structure}.

	\section{Proof of Theorem \ref{LP}}
	The power constraint (\ref{power}) can be easily obtained from the definition of $y_{q,s}$. Now we derive the expression of average delay (\ref{delay}). Consider the stationary probability of queue length being larger than $q$ and channel state being $s'$. On the one hand it can be characterized as follows:
	\begin{equation}
	\text{Pr}\{l>q,s=s'\}=\sum_{l=q+1}^{K}\mu_{l,s'}.
	\end{equation}
	On the other hand, it can be transferred from the previous slot whose queue length is either larger than $q$ or equal to $q$ while the node receives a new packet and does not transmit it. Therefore we have:
	\begin{equation}
	\text{Pr}\{l>q,s=s'\}=\sum_{s=1}^{S}P_{ss'}(\mu_{q,s}\theta+\sum_{l=q+1}^{K}\mu_{l,s}-y_{q,s}).
	\end{equation}
	Thus:
	\begin{equation}
	\sum_{s=1}^{S}P_{ss'}(\mu_{q,s}\theta+\sum_{l=q+1}^{K}\mu_{l,s}-y_{q,s})=\sum_{l=q+1}^{K}\mu_{l,s'}, \forall q,s.
	\end{equation}
	which is exactly (\ref{relationship}). From the above equation, we can obtain the transformation matrix $\mathbf{G}$. Take $s'$ from 1 to $S$ and sum the equations up, then we have:
	\begin{equation}
	\label{relationshipne}
	\sum_{x=1}^{S}y_{q,s}=\theta\sum_{s=1}^{S}\mu_{q,s}.
	\end{equation}
	By substituting the above equation into Little's Theorem, we have (\ref{delay}).
	Next we consider the steady-state constraint (\ref{packets}). Notice that the sum of stationary distribution should equal 1, so we have:
	\begin{equation}
	\sum_{q=0}^{K}\sum_{s=1}^{S}y_{q,s}=\theta\sum_{q=0}^{K}\sum_{s=1}^{S}\mu_{q,s}=\theta.
	\end{equation} 
	
	Since the communication channel is an ergodic Markov chain, we also needs to construct a constraint on the channel evolution. Consider the stationary distribution of the channel and we have:
	\begin{equation}
	\rho_{s}=\sum_{q=0}^{K}\mu_{q,s}.
	\end{equation} 
	Substitute (\ref{relation}) into the above equation and  we can obtain (\ref{channel}).
	Finally considering that transmission probability and stationary distribution should not surpass 1, we have (\ref{probability}) and (\ref{state}) respectively.
\end{appendices} 
\end{document}